\documentclass{aa}
\usepackage{graphicx}
\begin{document}
\title{NGC~5385, NGC~2664 and Collinder~21: three candidate 
Open Cluster Remnants
    \thanks{{}Based on observations
 carried out at Mt Ekar Observatory, Asiago, Italy},
     \thanks{Photometry is only available in electronic form at the CDS 
     via anonymous ftp to {\tt cdsarc.u-strasbg.fr (130.79.128.5)} or 
      via {\tt http://cdsweb.u-strasbg.fr/cgi-bin/qcat?J/A+A//}
             }
	}
\author{Sandro Villanova$^{1}$, Giovanni Carraro$^{1,2,3}$, 
Ra\'ul de la Fuente Marcos$^{4}$, and Ruggero Stagni $^{1}$}

\offprints{Sandro Villanova: villanova@pd.astro.it}

\institute{
             $^1$Dipartimento di Astronomia, Universit\`a di Padova,
                 Vicolo Osservatorio 2, I-35122 Padova, Italy\\
             $^2$Departamento de Astronom\'ia, Universidad de Chile, 
		 Casilla 36-D, Santiago, Chile\\
	     $^3$Astronomy Department, Yale University, P.O. Box 208101, 
		 New Haven, CT 06520-8101 USA\\
             $^4$Suffolk University Madrid Campus, C/ Vi\~na 3, E-28003, 
                 Madrid, Spain\\
           }

    \date{Received April 2004; accepted}

\abstract{We present CCD UBVI photometric and medium/high resolution 
          spectroscopic observations obtained in the field
          of the previously unstudied dissolving open cluster
          candidates NGC~5385, NGC~2664 and Collinder~21. Our 
          analysis stands on the discussion of star counts, photometry, 
          radial velocity distribution, and proper 
          motions available from the Tycho~2 catalogue. All the 
          three aggregates clearly emerge from the mean Galactic 
          field, but, regrettably, the close scrutiny of proper 
          motions and radial velocities reveals that we are not 
          facing any physical group. Instead, what we are looking 
          at are just chance alignments of a few bright unrelated stars.
Our analysis casts some doubt on the Bica et al. (2001) criterion
to look for Possible Open Cluster Remnants. It seems mandatory to
define a better criterion to adopt for further investigations.
}

\authorrunning{Villanova et al.}
\titlerunning{Possible Open Cluster Remnant Candidates}
\maketitle
 %

\section{Introduction}
 Bica et al. (2001) identified 34 neglected star clusters
 (see {\it http://obswww.unige.ch/webda/dissolving$\_$ocl.html}) 
 having relatively high galactic latitude ($|b| > 15^{o}$), being 
 poorly populated and appearing to be candidate objects to be
 experiencing the late stages of star cluster dynamical evolution. 
 For these objects the acronym POCR --Possible Open Cluster Remnant--  
 is used. In their study, Bica et al. (2001) basically select these
 candidates on the basis of star counts. All of them indeed clearly 
 emerge from the background. 
 \\
 The final residue of open star cluster evolution is often called an 
 open cluster remnant (OCR). These ghostly objects are characterized 
 by very low surface brightness and they are hardly distinguishable 
 against the background field stars (de la Fuente Marcos 1998). 
 They consist of a small number of coeval relatively massive stars,
 most of which are binaries, confined in a core, since due to the
 evolution the less massive star members evaporated from the cluster
 and merged with the Galactic disk field (de la Fuente Marcos 1997).\\
 Recently, these objects have started to receive some attention mainly
 because they play a fundamental role in our understanding of the
 subject of open cluster evolution and dissolution, and, ultimately, 
 on the origin of the field star population. Photometry, kinematics, 
 binary percentage and membership are basic data to constrain $N$-body 
 models of cluster dissolution, which aim to reconstruct 
 --for instance-- the Initial Mass Function (IMF) of Galactic clusters
 (e.g. de la Fuente Marcos 1997).\\
 On the other hand, they may be what still remains of old clusters, and 
 therefore once discovered they can allow us to improve on the statistics 
 of old open clusters in the Galactic disk. These objects are widely
 recognised to be of paramount importance for our understanding of
 the formation and early evolution of the Galactic disk (Carraro et al. 
 1998). Some POCR candidates have been studied recently.\\
 The best discussed case is doubtless NGC~6994 (M~73), studied by Bassino 
 et al. (2000), Carraro (2000) and Odenkirchen and Soubiran (2002). By 
 combining together photometry, astrometry and spectroscopy this object 
 turned out to be just a chance alignment of four stars.\\
 Pavani et al. (2001) discuss new photometry of NGC~1901 and NGC~1252, 
 bringing new evidence that these stellar groups are POCRs. However, proper 
 motions studies for NGC~1252 seem to point to the opposite conclusion 
 (Baumgardt 1998). On the other hand, the radial velocity survey conducted by
 Villanova et al. (2003a,b) confirms that NGC~1901 is a genuine POCR.\\
 Finally, on the basis of only star counts and photometry,
 Carraro (2002) discusses the cases of NGC~7772 and NGC~7036
 and Baume et al. (2003) the case of NGC~1663,
 favoring the possibility that these two objects be indeed OCRs and providing
 a list of possible candidate members for spectroscopic follow up.\\
 In this paper we present a photometric, astrometric and spectroscopic 
 investigation of three previously unstudied POCRs, namely 
 NGC~5385 and NGC~2664 and Collinder~21, with the goal
 to clarify their real nature. The analysis closely follows the 
 strategy earlier proposed by Odenkirchen and Soubiran (2002).\\

 \noindent
 The layout of this paper is as follows.\\
 In Sect.~2 we briefly present the observations and data reduction.
 Sect.~3 illustrates star counts analysis and spatial
 configurations. Sect.~4 is dedicated to the outcomes
 of the proper motion analysis, whereas in sect.~5 we discuss
 our spectroscopy. Our conclusions are presented in sect.6.

 \begin{table}
 \caption{Basic parameters of the observed objects.
 Coordinates are for J2000.0 equinox and have been 
taken from Dias et al. (2002)}
 \begin{tabular}{ccccc}
 \hline
 \hline
 \multicolumn{1}{c}{Name} &
 \multicolumn{1}{c}{$\alpha$}  &
 \multicolumn{1}{c}{$\delta$}  &
 \multicolumn{1}{c}{$l$} &
 \multicolumn{1}{c}{$b$} \\
 \hline
 & $hh:mm:ss$ & $^{o}$~:~$^{\prime}$~:~$^{\prime\prime}$ & $^{o}$
& $^{o}$ \\
 \hline
 NGC~5385       & 13:52:27 & +76:10:24 & 118.19 & +40.38\\
 NGC~2664       & 08:47:11 & +12:36:06 & 214.34 & +31.31\\     
 Collinder~21   & 01:50:11 & +27:04:00 & 138.73 & -33.99\\
 \hline\hline
 \end{tabular}
 \end{table}

 \section{Observations and Data Reduction}

\subsection{Photometry}
Observations were carried out with the AFOSC camera at the 
1.82~m Copernico telescope of Cima Ekar (Asiago, Italy), in the photometric
nights of December 17 and 18, 
2001. AFOSC samples a $8^\prime.14\times8^\prime.14$ field in a  
$1K\times 1K$ thinned CCD. The typical seeing was between 1.8  
and 2.3 arcsec.    
The basic data of the studied objects are summarized in Table~1, whereas
the details of the observations are listed in Table~2.
  
The data has been reduced by using the IRAF\footnote{IRAF  
is distributed by the National Optical Astronomy Observatories, 
which are operated by the Association of Universities for Research 
in Astronomy, Inc., under cooperative agreement with the National 
Science Foundation.} packages CCDRED, DAOPHOT, and PHOTCAL. 
The calibration equations obtained by observing Landolt(1992) 
SA~93, PG~1047+003, PG~2331+055 and PG~0231+051 fields along both the nights,
are:
  
         \begin{eqnarray}  
 \nonumber 
 u \! &=& \! U + 3.520\pm0.042 + (0.099\pm0.030)(U\!-\!B) + 0.58\,X \\  
 \nonumber 
 b \! &=& \! B + 1.407\pm0.012 - (0.004\pm0.017)(B\!-\!V) + 0.29\,X \\  
 \nonumber 
 v \! &=& \! V + 0.752\pm0.009 + (0.036\pm0.012)(B\!-\!V) + 0.16\,X \\  
 \nonumber 
 i \! &=& \! I + 1.619\pm0.017 - (0.011\pm0.015)(V\!-\!I) + 0.08\,X \\ 
         \label{eq_calib} 
         \end{eqnarray}
 \noindent 
where $UBVI$ are standard magnitudes, $ubvi$ are the instrumental  
ones, and $X$ is the airmass. The standard stars in these fields
provide a very good color coverage, being $-0.329 \leq (B-V) \leq 1.448$.
For the extinction coefficients, 
we assumed the typical values for the Asiago Observatory
(Desidera et al. 2001).
Photometric global errors have been estimated following Patat
and Carraro (2001). For the $V$ filter,  
they amount to 0.02, 0.04 and 0.08 at $V\approx$
12.0, 16.0 and 20.0, respectively. \\
The photometry for the most obvious candidates is reported in
Tables~3 to 5, whereas all the photometric catalogues are available upon
request to the authors.

 \begin{table} 
 \tabcolsep 0.30truecm 
 \caption{Journal of photometric observations in the field of 
 Collinder~21, NGC~5385,  and NGC~2664 
 and standard star fields (December 17-18, 2001).} 
 \begin{tabular}{cccc} 
 \hline 
 \multicolumn{1}{c}{Field}    & 
 \multicolumn{1}{c}{Filter}    & 
 \multicolumn{1}{c}{Time integration}& 
 \multicolumn{1}{c}{Seeing}         \\ 
       &        & (sec)     & ($\prime\prime$)\\ 
   
 \hline 
  Collinder~21   &     &              &      \\ 
                 & $U$ &  15,60       &  2.1 \\ 
                 & $B$ &  2,5,15      &  2.0 \\ 
                 & $V$ &  1,3,10,10   &  2.1 \\ 
                 & $I$ &  1,8         &  2.0 \\ 
  SA~93          &     &              &      \\ 
                 & $U$ &  120         &  1.9 \\
                 & $B$ &  60,60       &  2.0 \\ 
                 & $V$ &  30,30,30    &  2.0 \\ 
                 & $I$ &  30,30       &  2.0 \\ 
  NGC~5385       &     &              &      \\ 
                 & $U$ &  240         &  2.2 \\ 
                 & $B$ &  60,60       &  2.1 \\ 
                 & $V$ &  15,15       &  2.3 \\ 
                 & $I$ &  15,15       &  2.3 \\
 PG~1047+003     &     &              &      \\ 
                 & $U$ &  120         &  1.9 \\
                 & $B$ &  60,60       &  2.0 \\ 
                 & $V$ &  30,30       &  2.0 \\ 
                 & $I$ &  30,30       &  2.0 \\ 
 PG~0231+051     &     &              &      \\ 
                 & $U$ &  120         &  1.9 \\
                 & $B$ &  60,60       &  2.0 \\ 
                 & $V$ &  30,30       &  2.0 \\ 
                 & $I$ &  30,30       &  2.0 \\ 
 PG~2331+055     &     &              &      \\ 
                 & $U$ &  120         &  1.9 \\
                 & $B$ &  60,60       &  2.0 \\ 
                 & $V$ &  30,30       &  2.0 \\ 
                 & $I$ &  30,30       &  2.1 \\ 
  NGC~2664       &     &              &      \\ 
                 & $U$ &  15,240      &  1.8 \\ 
                 & $B$ &  1,10,60     &  1.7 \\ 
                 & $V$ &  1,5,30,30   &  1.8 \\ 
                 & $I$ &  1,5,30      &  1.8 \\
 \hline 
 \end{tabular} 
 \end{table}

\subsection{Spectroscopy}
$\hspace{0.5cm}$
Medium ($R \approx 3,600$) and high resolution  ($R \approx 20,000$)
spectra in the field of NGC~5385, NGC~2664 and Collinder~21
have been obtained
by using AFOSC in Echelle mode (in the grism
$\#9$ and $\#10$ combination) and the REOSC Echelle
Spectrograph onboard the 1.82~m telescope of Asiago Astronomical 
Observatory.\\ 
The Echelle 
spectrograph works with a Thomson 1024$\times$1024 CCD and the 
allowed wavelength coverage is approximately $4140 - 6840$ \AA. 
Details on this 
instrument are given in Munari \& Zwitter (1994) and at the Asiago
Observatory Home 
page\footnote{http://www.pd.astro.it/Asiago/2000/2300/2320.html}. \\
The exposures times were 45 minutes for the all stars. In order to improve the 
signal to noise ratio, two exposures were taken for each star reaching at the 
end $S/N$ values up to 150. The data have been reduced with the IRAF package 
ECHELLE using thorium lamp spectra for wavelength calibration purposes. By 
comparing final known sky lines positions along the spectra we obtained an
error measurement of about 0.01 \AA.\\
The AFOSC in Echelle mode spectra cover the wavelength range $4100 - 6800$ \AA,
and for typical exposure times of 30 to 60 minutes we obtained
$S/N$ values up to 300, which allows us to derive radial velocity with error
almost always lower than 8 km/s.
In this case, the data has been reduced with the IRAF package for 
one-dimensional spectroscopy {\em CTIOSLIT} and by using the Hg-Cd lamp spectra 
for wavelength calibration purposes.
Together with radial velocities we provide also spectral classifications,
which have been derived as described in Villanova et al. (2004).\\
In Tables 6 to 8 we report some details of the spectroscopic observations.\\

    \begin{figure}
    \centering
    \includegraphics[width=9cm]{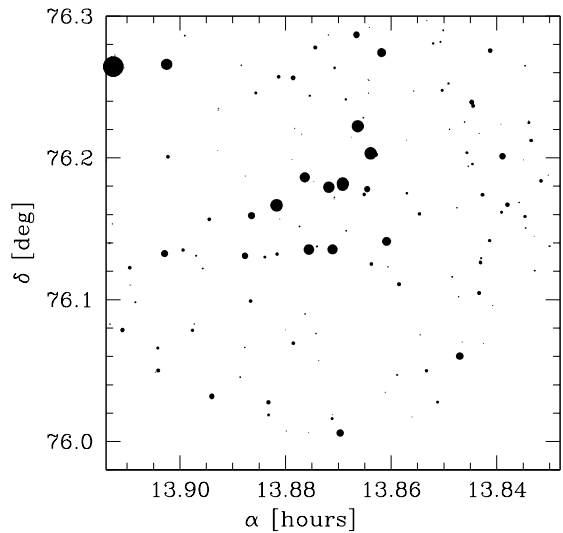}
    \caption{Distribution of stars from USNO-B1 catalogue
   in a $20^{\prime} \times 20 ^{\prime}$ field around NGC~5385.
   The size of the dots are proportional to the magnitude
   of the stars.}
     \end{figure}

    \begin{figure}
    \centering
   \includegraphics[height=11cm,width=9cm]{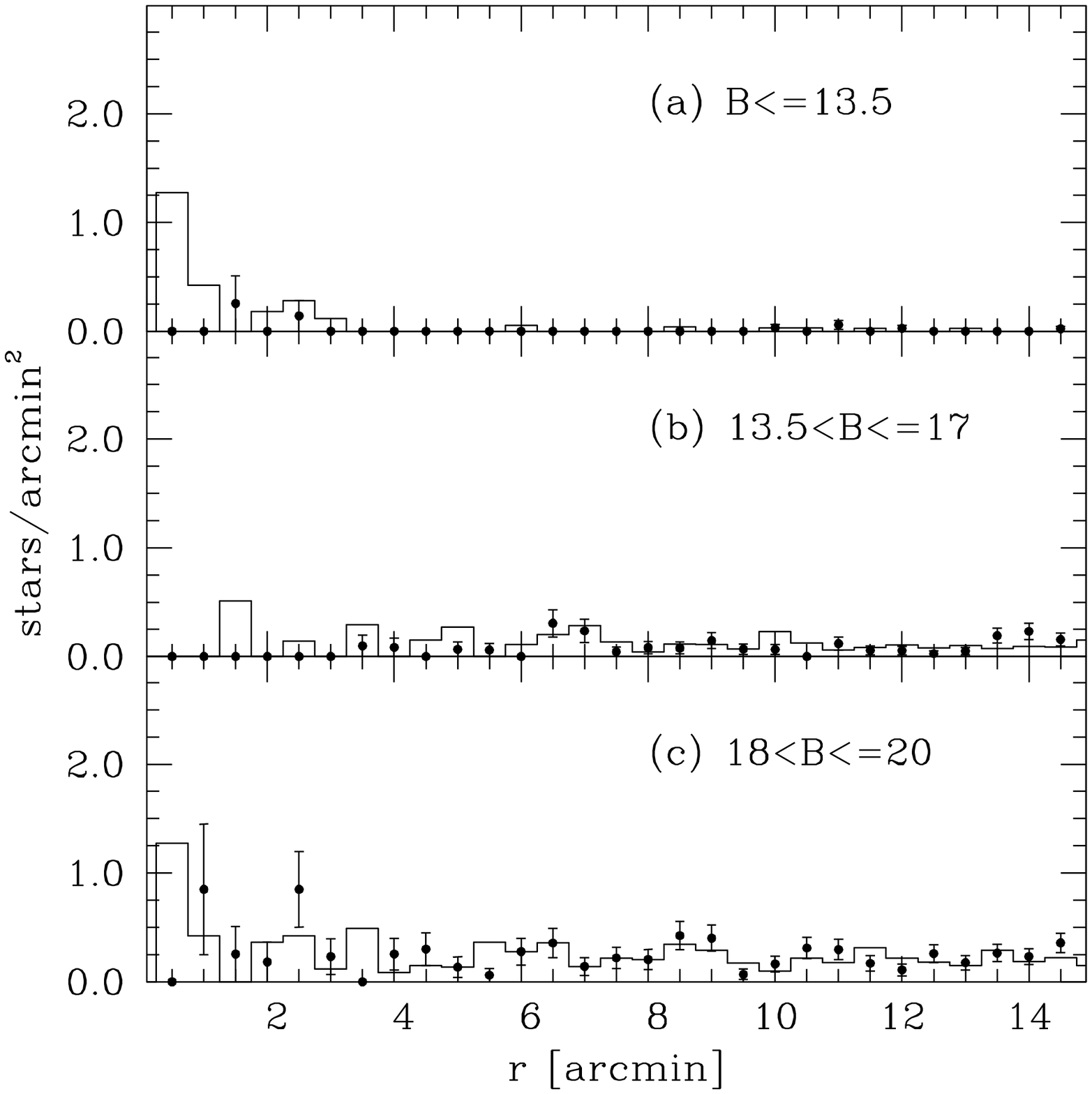}
    \caption{Histogram of the surface density of the stars shown
   in Fig.~1, measured by counts in $0^{\prime}.5$ wide annuli centered on
$\alpha=13^h52^m19^s$, $\delta=+76^{o}10^{\prime}56^{\prime\prime}$. 
Dots with error bars indicate the mean surface density
         of stars in a surrounding field centered on 
$\alpha=13^h54^m10^s$, $\delta=+76^{o}49^{\prime}30^{\prime\prime}$ 
, and the expected
 1$\sigma$ range of statistical fluctuations around this mean for each bin.}
     \end{figure}

\section{Star counts, surface density and spatial distribution}
POCRs are identified by Bica et al. (2001) as star overdensities at 
relatively high galactic latitude. 
For star count purposes we are going to use 
USNO-B1 (Monet et al. 2003) catalogue for NGC~5385
and UCAC2 (Zacharias et al. 2003) catalogue for NGC~2664 and Collinder~21,
because UCAC2 does not extend to NGC~5385 declination.
In the following, we are comparing star counts inside the OCR region
with star counts in a off-set field, which is aimed to represent the mean 
Galactic disk field.\\ 
A crucial role in star counts is played by
binaries. Anticipating the results of Section~5, we find that in the OCRs
under investigation the binary fraction ranges from 18$\%$ to  25$\%$, a value which
is not much different from the estimated typical
percentage in the Galactic disk field population, which in the the proper $F-K$ spectral
range amounts to about 14$\%$ (Halbwachs et al 2003). If one considers also $M$ type stars,
this percentage would increase.\\
Therefore, we do not expect that
binary stars significantly affect star counts.

\subsection{NGC~5385}
Fig.~1 shows a finding chart in a  $20^{\prime} \times 20^{\prime}$
region around NGC~5385 down to B = 21.5
according to USNO-B1. The center of this image, which we also
consider the center of NGC~5385, has been derived as the mean
of the position of the 9 most obvious members of the aggregate.
Indeed this object seems to consist of 9 stars with magnitudes
11$\leq B \leq$ 13 in a field of about $3^{\prime}.5 \times 3^{\prime}.5$
The aggregate appears rather sparse, being the mean mutual
angular separation larger than  $30^{\prime\prime}$.
Out of the central group, there is a clear deficiency of similar
brightness stars, which appear at least 1 mag fainter on the average, 
if we exclude some obvious case northward NGC~5385.\\
In Fig.~2 we show the surface density of stars of different magnitudes
as a function of angular distance from the center of NGC~5385, (based
on USNO-B1). The central group of bright stars produces a significant
local enhancement in the surface density of stars with magnitudes
$B \leq$ 13.5 (see panel (a)).
Fainter stars in the magnitude bins $13.5 \leq B \leq 17.0$ (panel (b))
$B \geq 18.0$ (panel (c))
do not show any signs of a spatial concentration, star counts 
in the cluster area always agree within the uncertainties with
the surrounding field ones. In fact in the lower panel the overdensity
in the central part lies quite close to the noise level to be
considered significant.\\
In conclusion only the brightest stars indicate the existence
of a cluster, and we are here going to look for membership in this
magnitude range. This of course does not exclude that also fainter
stars could be members, in the case the cluster turns out
to be a physical one.

\subsection{NGC~2664}
Fig.~3 shows the distribution of stars in a field of 
$30^{\prime} \times 30^{\prime}$ around NGC~2664 down to
a magnitude UCmag (between V and R) of $\approx$ 16.3 according to UCAC2. 
This figure reveals that NGC~2664 is defined by 4 stars
with magnitude $10.5 \leq UCmag \leq 11.5$ confined in $2^{\prime}.5
\times 2^{\prime}.5$arcmin around the nominal center, taken as
the mean of the 4 most obvious members positions.
In the immediate surroundings 
there is a clear lack of bright stars, although outward about $3^{\prime}$ 
from the cluster center the field is quite rich in bright stars, and
a few other concentrations of the same kind as NGC~2664 
seem to be present.\\
In Fig.~4 we show the surface density of stars of different magnitudes
as a function of angular distance from the center of NGC~2664, (based
on UCAC2 as well). The central group of bright stars produces a small but
sizeable
local enhancement in the surface density of stars with magnitudes
$ UCmag\leq$ 13.5 (see panel (a)).
The same seems to occur for the magnitude bins $13.0 \leq  UCmag \leq 16.0$ 
(panel (b)), whereas fainter stars, in the magnitude bin
$UCmag \geq 16.0$ (panel (c))
do not show any clear sign of a spatial concentration. At all distances
in fact the cluster density profile is comparable with the field, and only
at a distance of 3 arcmin does there seem to be a significant enhancement.\\
Therefore, the asterism NGC~2664 seems to be mainly identified by four
bright stars and a few fainter stars in its immediate vicinity.

    \begin{figure}
    \centering
    \includegraphics[width=9cm]{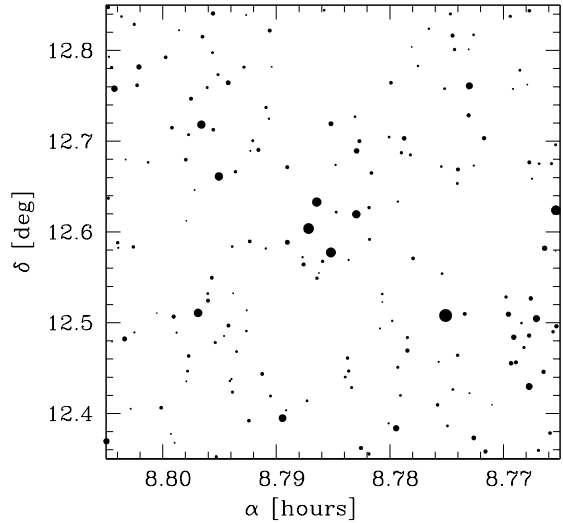}
    \caption{Distribution of stars from UCAC2 catalogue
   in a $30^{\prime} \times 30^{\prime}$ field around NGC~2664.
   The size of the dots are proportional to the magnitude
   of the stars.}
     \end{figure}

    \begin{figure}
    \centering
   \includegraphics[height=11cm,width=9cm]{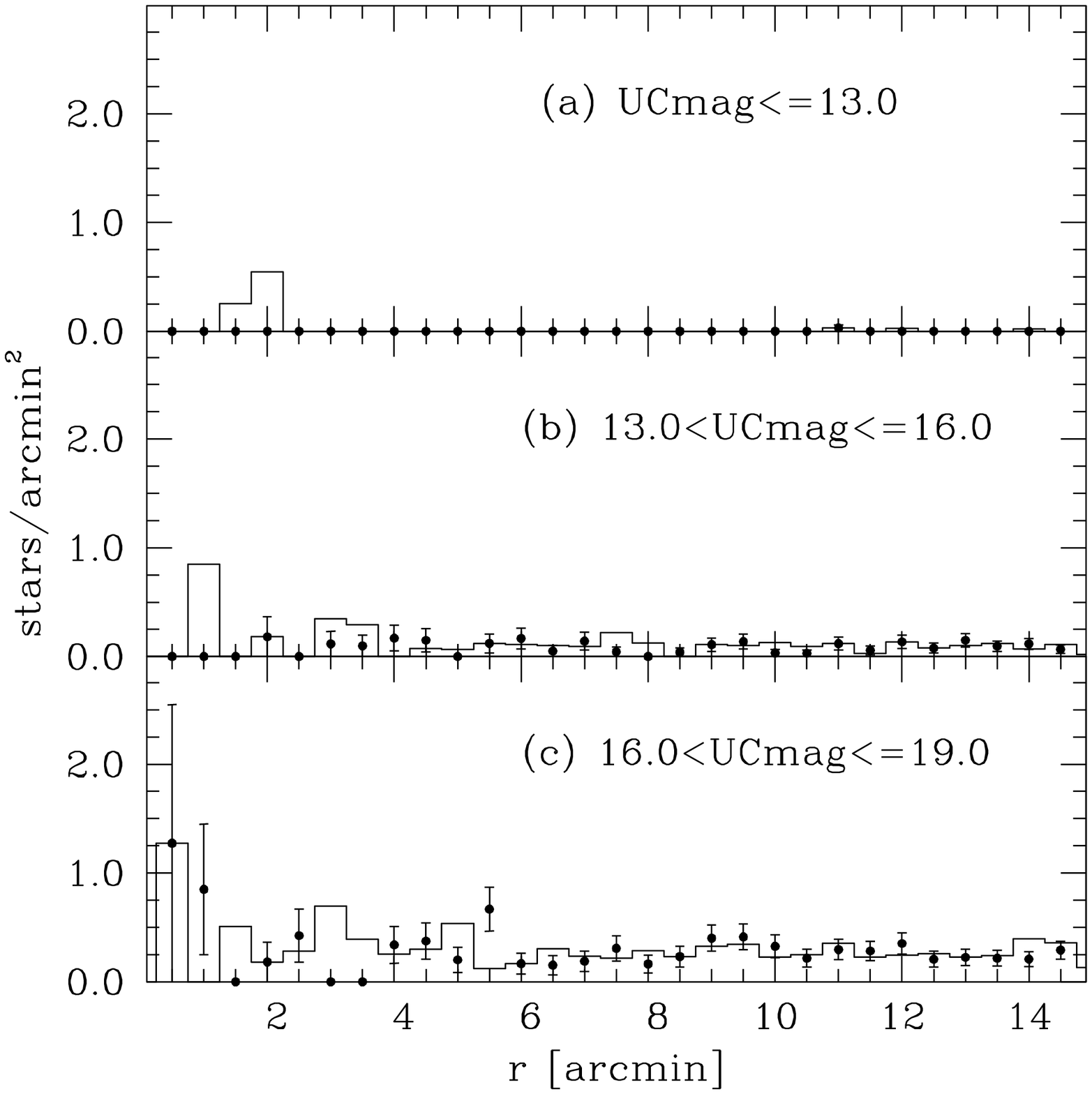}
    \caption{Histogram of the surface density of the stars shown
   in Fig.~3, measured by counts in $0^{\prime}.5$ wide annuli centered on
$\alpha=08^h47^m07^s$, $\delta=+12^{o}36^{\prime}30^{\prime\prime}$.  
Dots with error bars indicate the mean surface density
         of stars in a surrounding field centered on 
$\alpha=08^h53^m05^s$, $\delta=+12^{o}59^{\prime}50^{\prime\prime}$, 
and the expected
 1$\sigma$ range of statistical fluctuations around this mean for each bin.}
     \end{figure}

\subsection{Collinder~21}
Fig.~5 shows the distribution of stars in a field of 
$30^{\prime} \times 30^{\prime}$ around Collinder~21 down to
a magnitude of $\approx$ 15.3 according to UCAC2. 
The appearance of Collinder~21 (also designated as OCL~371
and C0147+270) on the sky is very impressive.
It consists of about 10 stars distributed in a ring-like structure
in a very poorly populated field, which makes this structure to
emerge very sharply. There are two well known binary stars in this
circlet: the visual binary system BD+26~305AB and the
binary HD~11142, resolved by speckle interferometry,
which presents a separation 
of $0^{\prime\prime}.56$ between its components. All the other
stars have much larger angular separation.\\

    \begin{figure}
    \centering
    \includegraphics[width=9cm]{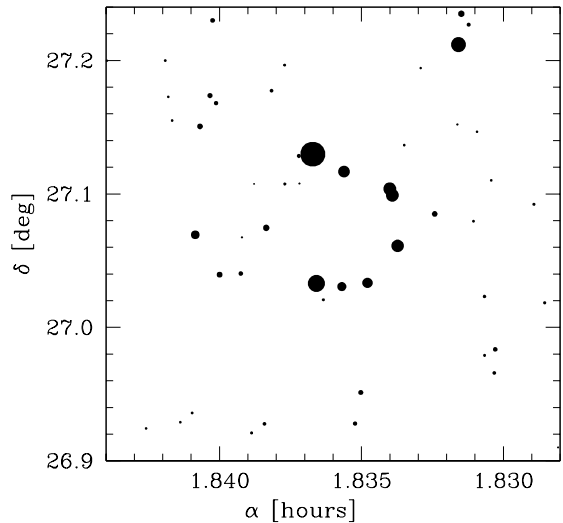}
    \caption{Distribution of stars from UCAC2 catalogue
   in a $30^{\prime} \times 30^{\prime}$ field around Collinder~21.
   The size of the dots are proportional to the magnitude
   of the stars.}
     \end{figure}

    \begin{figure}
    \centering
   \includegraphics[height=11cm,width=9cm]{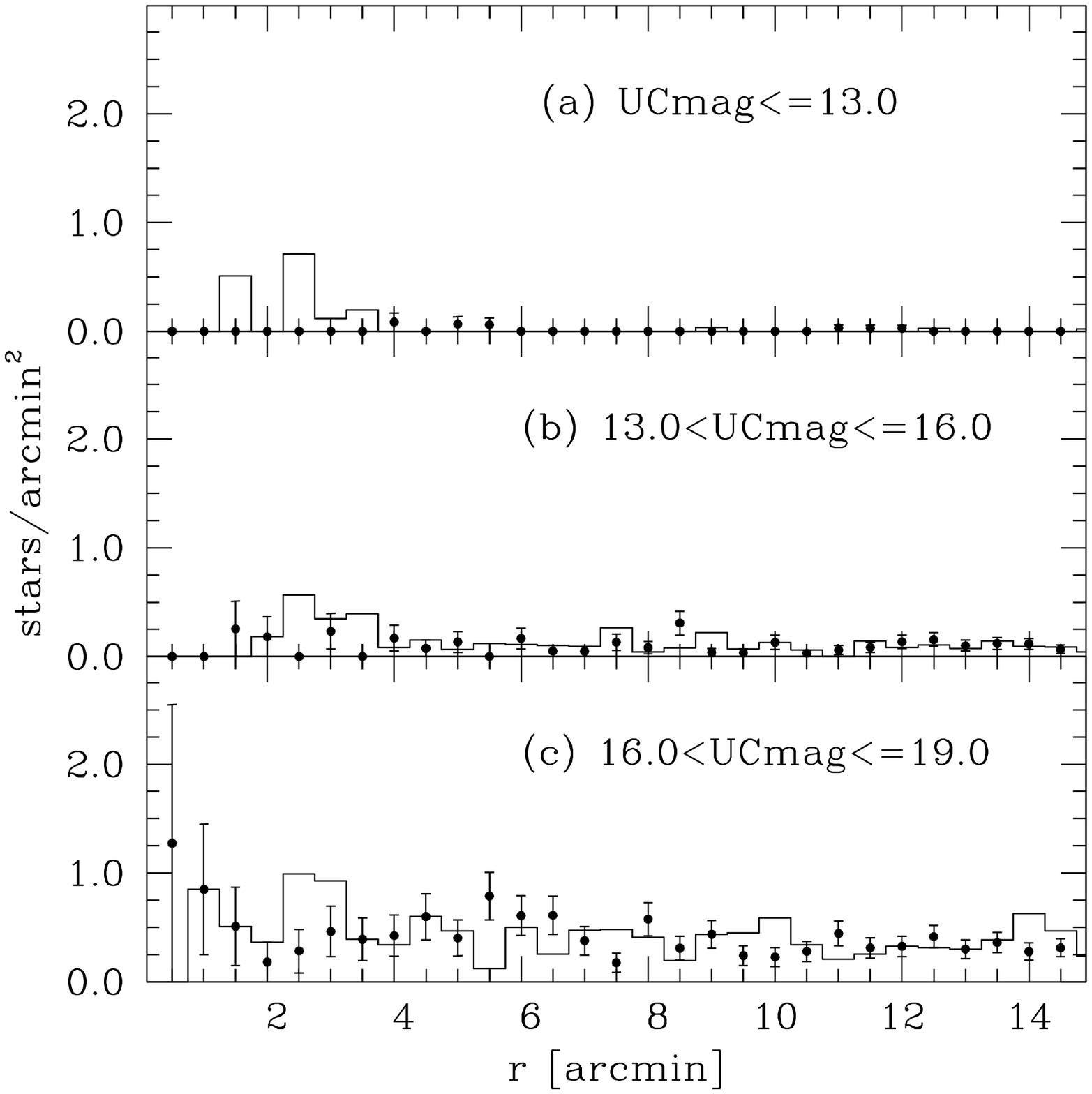}
    \caption{Histogram of the surface density of the stars shown
   in Fig.~5, measured by counts in $0^{\prime}.5$ wide annuli centered on
$\alpha=01^h50^m11^s$, $\delta=+27^{o}04^{\prime}26^{\prime\prime}$.  
Dots with error bars indicate the mean surface density
    of stars in a surrounding field centered on 
$\alpha=01^h54^m10^s$, $\delta=+27^{o}44^{\prime}00^{\prime\prime}$, 
and the expected
 1$\sigma$ range of statistical fluctuations around this mean for each bin.}
     \end{figure}

\begin{figure}
\includegraphics[width=9cm]{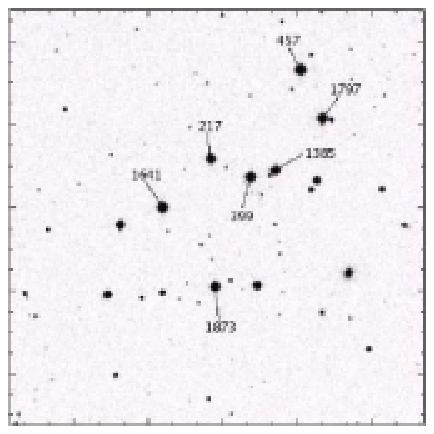}
\caption{Identification map ($8^{\prime}.14 \times 8^{\prime}.14$), taken from
DSS-2)
for member candidates of NGC~5385 having proper motions measurements from
Tycho~2. North is up, East on the left. The labels give the star numbers
assigned in
the Tycho-2 catalog (in the sense TYC 4558$-<...>$).Data for these stars
are given in Tables~3 and 6.}
\end{figure}

In Fig.~6 we show the surface density of stars of different magnitudes
as a function of angular distance from the center of Collinder~21, based
on USNO-B1. The bright star belonging to the circlet produce a significant
local enhancement in the surface density of stars with magnitudes
 $ UCmag \leq$ 13.0 (see panel (a)) at a distance corresponding to the
radius of the circlet.
Fainter stars in the magnitude bins $13.0 \leq UCmag \leq 16.0$ (panel (b))
$UCmag \geq 16.0$ (panel (c))
do not show any signs of a spatial concentration since
star counts 
in the cluster area always agree within the uncertainties with
the surrounding fields ones. \\
The only exception is again at a distance of about 3 arcmin, where
the enhancement is alway significant at 2$\sigma$ level.
In conclusion star counts suggests that a  clear concentration
exist in the region of Collinder~21.\\

\noindent
At the end of this section, we would like to stress that the existence
of some stars overdensity is a necessary condition for a physical cluster
to exist, but not a sufficient one. We repeat here the kind of analysis
proposed by Odenkirchen and Soubiran (2002), to show what actually number
statistics can tell us about these 3 objects.\\
We use their Equation (1):

\begin{equation}
p(n) = \frac{1}{n!} (\Sigma \pi \Theta^2)^n exp (-\Sigma \pi \Theta^2)
\end{equation}

\noindent
which gives us the probability of finding $n$ neighbours in a circular
field of radius $\Theta$ around an arbitrarily selected star,
for a random distribution of stars with mean surface density
$\Sigma$.\\
The statistics of star counts in the comparison fields
 for NGC~5385, NGC~2664 and
Collinder~21 provide in the magnitude range $10.0 \leq mag \leq 13.0$
a mean surface density of 20, 16 and 64 stars/deg$^2$, respectively.\\
In the case of NGC~2664, we are looking at the probability
to find 3 or more neighbours within a radius of 2.5 arcmin.
As a consequence the probability of finding 3  or more neighbours turns out
to be lower than 1.2 $\times$ 10$^{-4}$.
As for NGC~5385, we must look for the probability to find at least 8
neighbours within about 3 arcmin, which amounts to 
less than 1.1 $\times$ 10$^{-11}$.
Finally, in the case of Collinder~21, we need at least 5 neighbours
within about 3 arcmin, and the corresponding probability
results lower than 4.4 $\times$ 10$^{-4}$.\\
These numbers are saying to us that random configurations like those
ones we are facing are rare. 
As a consequence, in all the  three cases (and especially for
 NGC~5385) 
we cannot exclude these are physical groupings,
and therefore only proper motions
and radial velocities can settle the question of their real nature.

\begin{figure*}
\includegraphics[width=18cm]{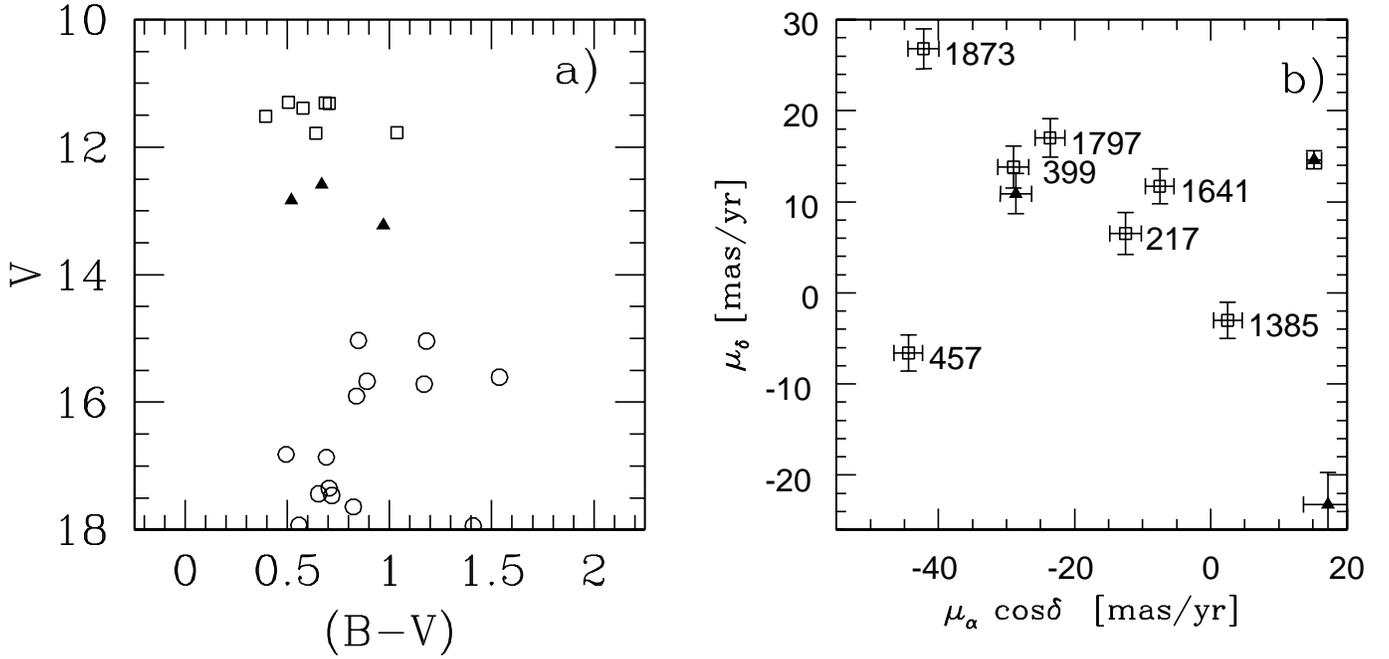}
\caption{Photometry and proper motions in the region of NGC~5385.
{\bf a)} Color-Magnitude diagram. Open squares are Tycho~2 stars 
having both proper motion and radial velocity measurements
(see panel b) and Table~6), while
filled triangles are stars having only proper motion
measurements. {\bf b)} Vector point
plot of Tycho~2 proper motions and proper motion errors. 
Symbols are as in panel {\bf a)}. As for the identification,
we are plotting here Tycho~2 numbering. See text for more details.}
\end{figure*}

\begin{table*}
\tabcolsep 0.12cm
\caption{Photometry and proper motion of the most obvious 
candidate members in the field of NGC~5385.}
\begin{tabular}{ccccccccc}
\hline
\hline
\multicolumn{1}{c}{ID} &
\multicolumn{1}{c}{TYC 4558-} &
\multicolumn{1}{c}{$\alpha (J2000.0)$}&
\multicolumn{1}{c}{$\delta (J2000.0)$}&
\multicolumn{1}{c}{$V$}  &
\multicolumn{1}{c}{$(B-V)$} &
\multicolumn{1}{c}{$(U-B)$}  &
\multicolumn{1}{c}{$\mu_\alpha cos \delta$ [mas/yr]} &
\multicolumn{1}{c}{$\mu_\delta$ [mas/yr]}  \\
\hline
   1& 1641& 13:52:54.104& +76:09:59.63&  11.29& 0.50& 0.18&   -7.5$\pm$1.9&  11.7$\pm$2.1\\
   2& 1385& 13:52:08.976& +76:10:55.12&  11.52& 0.39& 0.14&    2.5$\pm$2.0&  -3.0$\pm$2.1\\
   3&  457& 13:51:58.757& +76:13:20.41&  11.39& 0.58& 0.32&  -44.6$\pm$2.0&  -6.6$\pm$2.1\\
   4&  399& 13:52:18.535& +76:10:45.58&  11.31& 0.68& 0.39&  -29.0$\pm$2.3&  13.8$\pm$2.3\\
   5& 1797& 13:51:49.975& +76:12:11.67&  11.32& 0.72& 0.56&  -23.6$\pm$2.1&  17.0$\pm$2.2\\
   6& 1873& 13:52:32.078& +76:08:07.41&  11.78& 0.64& 0.33&  -42.2$\pm$2.2&  26.8$\pm$2.3\\
   7&  217& 13:52:34.850& +76:11:10.84&  11.77& 1.04& 1.03&  -12.5$\pm$2.3&   6.5$\pm$2.4\\
\hline
\hline
\end{tabular} 
\end{table*}
 
\begin{table*}
\tabcolsep 0.12cm
\caption{Photometry and proper motion of the most obvious
candidate members in the field of
NGC~2664 .}
\begin{tabular}{ccccccccccc}
\hline
\hline
\multicolumn{1}{c}{ID} &
\multicolumn{1}{c}{TYC 816-} &
\multicolumn{1}{c}{Name} &
\multicolumn{1}{c}{$\alpha (J2000.0)$}&
\multicolumn{1}{c}{$\delta (J2000.0)$}&
\multicolumn{1}{c}{$V$}  &
\multicolumn{1}{c}{$(B-V)$} &
\multicolumn{1}{c}{$(U-B)$}  &
\multicolumn{1}{c}{$(V-I)$} &
\multicolumn{1}{c}{$\mu_\alpha cos \delta$ [mas/yr]} &
\multicolumn{1}{c}{$\mu_\delta$ [mas/yr]} \\
\hline
   1& 2354& BD+13~1898 & 08:47:11.173& +12:37:58.63&  10.75& 0.43& 0.19& 0.69& -6.4$\pm$1.9&   5.6$\pm$2.1\\
   2& 1826&            & 08:46:58.629& +12:37:10.47&  11.11& 0.37& 0.29& 0.61& -8.3$\pm$2.0&  -4.7$\pm$2.1\\
   3& 1890&            & 08:47:13.795& +12:36:13.50&  10.99& 1.03& 1.05& 1.13&  1.5$\pm$2.0& -26.0$\pm$2.1\\
   4& 2400& BD+13~1899 & 08:47:06.699& +12:34:38.76&  11.41& 1.11& 0.99& 1.15& -2.6$\pm$2.3&   3.2$\pm$2.3\\
\hline
\hline
\end{tabular}
\end{table*}

\begin{table*}
\tabcolsep 0.08cm
\caption{Photometry and proper motion of the most obvious candidate members in
the field
of Collinder~21.}
\begin{tabular}{ccccccccccc}
\hline
\hline
\multicolumn{1}{c}{ID} &
\multicolumn{1}{c}{TYC 1759-} &
\multicolumn{1}{c}{Name} &
\multicolumn{1}{c}{$\alpha (J2000.0)$}&
\multicolumn{1}{c}{$\delta (J2000.0)$}&
\multicolumn{1}{c}{$V$}  &
\multicolumn{1}{c}{$(B-V)$} &
\multicolumn{1}{c}{$(U-B)$}  &
\multicolumn{1}{c}{$(V-I)$} &
\multicolumn{1}{c}{$\mu_\alpha cos \delta$ [mas/yr]} &
\multicolumn{1}{c}{$\mu_\delta$ [mas/yr]} \\
\hline
   1&  450& BD+26~307 & 01:50:11.56& +27:01:59.0&   8.20& 0.45&     &      & 56.5$\pm$1.5& -33.5$\pm$0.9\\
   2& 1472& BD+26~304 & 01:50:01.41& +27:03:40.0&   9.80& 0.61&     & 0.82 &  5.1$\pm$1.2& -11.0$\pm$1.2\\
   3& 1025& BD+26~306 & 01:50:12.17& +27:07:47.1&   9.69& 1.75& 1.65&      & -8.7$\pm$1.4&  -1.1$\pm$1.4\\
   4&  468&           & 01:50:08.22& +27:07:00.2&  11.32& 1.39& 1.50& 1.46 & -0.7$\pm$2.3&  -3.3$\pm$2.3\\
   5&  994&           & 01:50:13.40& +27:05:57.5&  11.17& 0.74& 0.44& 0.95 & 48.7$\pm$1.9& -11.5$\pm$1.9\\
   6&  462&           & 01:50:27.06& +27:04:09.7&  11.32& 0.59& 0.26& 0.83 & -6.4$\pm$1.9&  -8.4$\pm$1.9\\
   7& 1333&           & 01:50:08.53& +27:01:49.4&  11.61& 0.89& 0.69& 1.07 & -6.9$\pm$1.2& +16.7$\pm$0.9\\
   8&  266&           & 01:50:05.23& +27:02:00.0&  11.21& 0.98& 0.79& 1.14 & 24.1$\pm$1.9&   4.3$\pm$1.9\\
   9&  866& BD+26~305b& 01:50:02.41& +27:06:13.8&  10.09& 0.85& 0.58& 1.04 & -3.9$\pm$1.3&  -6.7$\pm$1.2\\
  10& 1114& BD+26~305a& 01:50:02.08& +27:05:56.6&  10.36& 1.08& 1.08& 1.18 & -3.6$\pm$1.8&  -5.0$\pm$1.8\\
  11& 1749&           & 01:50:18.14& +27:04:28.9&  12.34& 0.56& 0.27& 0.78 & 11.9$\pm$2.3&  -5.5$\pm$2.3\\
  12&  272&           & 01:50:23.97& +27:02:23.1&  12.81& 0.65& 0.27& 0.86 &  1.9$\pm$1.3&  -8.4$\pm$2.9\\
\hline
\hline
\noalign{\smallskip}\noalign{Note: star $\#$1 was saturated in our photometry,
and magnitude and color have been taken from Tycho-2. Stars $\#$7 and  $\#$12 have been taken from UCAC2 catalog.}
\end{tabular}
\end{table*}

\section{Proper motion analysis}
Important information on the kinematics of the luminous stars in and around
our targets can be derived from the proper motions available in the 
Tycho-2 catalogue (H{\o}g et al. 2000).  
We decided to opt for this catalogue since it provides
homogeneous proper motions data for all the targeted stars. The recently
released UCAC2 catalogue in fact does not provide 
data for Collinder~21.\\
In Figs. 7, 9 and 11 we show the finding
charts of our targets, in a field which roughly corresponds to our photometric
survey, and indicate there the stars for which we have proper motion
measurements 
at our disposal. Moreover, Tables 3 to 5 list proper motion values,
magnitudes and colors for the most obvious candidate members of the aggregates
under investigation.

\begin{figure}
\includegraphics[width=9cm]{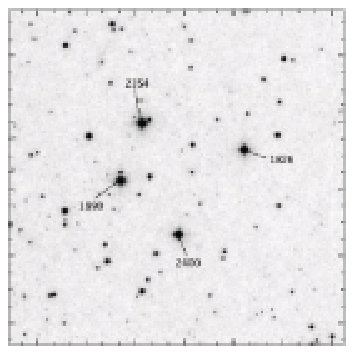}
\caption{Identification map ($8^{\prime}.14 \times 8^{\prime}.14$, taken from
DSS-2)
for member candidates of NGC~2664 having proper motions measurements from
Tycho~2. North is up, East on the left. The labels give the star numbers
assigned in
the Tycho-2 catalog (in the sense TYC 816$-<...>$.
Data for these stars are given
in Tables~4 and 7.}
\end{figure}

\begin{figure*}
\includegraphics[width=18cm]{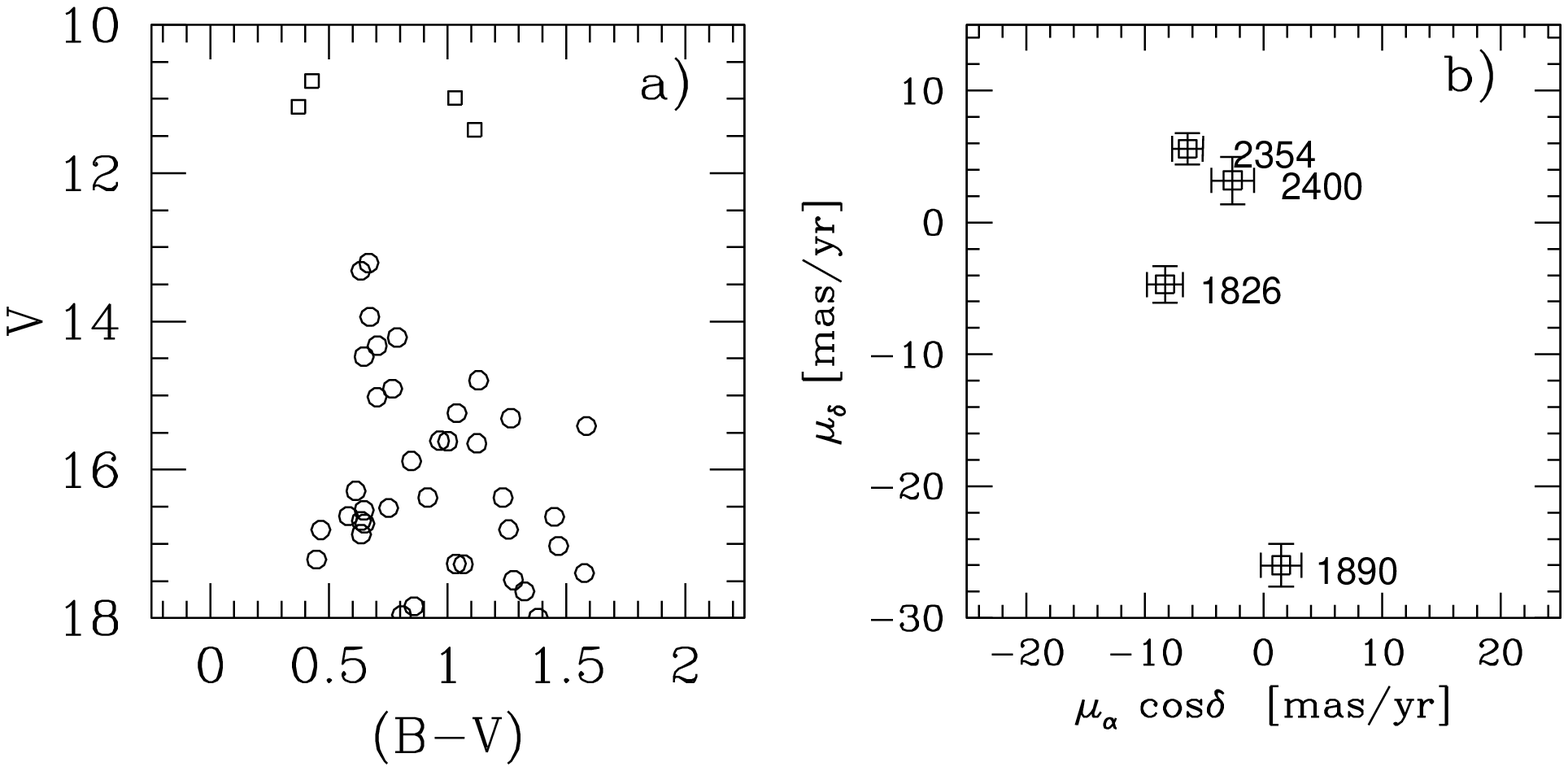}
\caption{Photometry and proper motions in the region of NGC~2664.
{\bf a)} Color-Magnitude diagram. Open squares are Tycho~2 stars 
having both proper motion and radial velocity measurements
(see panel b) and Table~7), while
filled triangles are stars having only proper motion
measurements. {\bf b)} Vector point
plot of Tycho~2 proper motions and proper motion errors. 
Symbols are as in panel {\bf a)}.As for the identification,
we are plotting here Tycho~2 numbering. See text for more details.}
\end{figure*}

  \begin{table*}
 \tabcolsep 0.40truecm 
  \vspace{1.cm}
  \caption[]{Spectroscopic results for NGC~5385}
  \label{tab2}
  \begin{tabular}{cccccccccc}
  \hline
  \noalign{\smallskip}
  \noalign{\smallskip}
  ID & TYC & Date of& S/N &Resolution &Julian Date  & Rad. vel
& Sp. Type & [Fe/H] & (m-M)$_V$\\
 &  4558-   & Observation &  &   & & [km/s] &  & [dex] & [mag] \\
  \noalign{\smallskip}
  \hline
  \noalign{\smallskip}
1 & 1641  & 12/07/2002 & 120& 20000& 2452467.4011& $-10.5\pm
1.2$ & F6 V& +0.2$\pm$0.3&7.49\\
&  & 15/02/2003 &    & 20000& 2452685.6788& $-7.5\pm 1.0$
& \\
&  & 12/05/2003 &    & 20000& 2452772.3750& $-6.9\pm 0.3$
& \\
  \noalign{\smallskip}
2 & 1385  & 12/07/2002 &110 & 20000& 2452468.3527  &
$-61.8\pm2.3$ & F4 V& +0.4$\pm$0.3  & 8.07\\
&  & 15/02/2003 &    & 20000& 2452685.5271  & $-62.5\pm1.0$
& \\
&  & 12/05/2003 &    & 20000& 2452772.4526  & $-62.4\pm0.3$
& \\
  \noalign{\smallskip}
3 & 457  & 12/07/2002 &120 & 20000& 2452468.4014  & $-1.9\pm1.3$
& F8 V& +0.9$\pm$0.4 & 7.19\\
&  & 15/02/2003 &    & 20000& 2452685.6436  & $-1.1\pm1.0$
& \\
&  & 12/05/2003 &    & 20000& 2452772.5267  & $-1.5\pm0.3$
& \\
  \noalign{\smallskip}
4 & 399  & 12/07/2002 & 120& 20000& 2452467.4495  & $2.5\pm1.8$
& G2 V& +0.4$\pm$0.3 & 6.51\\
&  & 15/02/2003 &    & 20000& 2452685.5377  & $1.1\pm1.0$
& \\
&  & 12/05/2003 &    & 20000& 2452772.4272  & $4.9\pm0.3$
& \\
  \noalign{\smallskip}
5 & 1797  & 12/07/2002 &120 & 20000& 2452468.4493  &
$-12.0\pm0.7$ &  G6 IV& +0.3$\pm$0.3 & 8.17\\
&  & 15/02/2003 &    & 20000& 2452685.6092  & $-11.3\pm1.0$
& \\
&  & 12/05/2003 &    & 20000& 2452772.5043  & $-12.3\pm0.3$
& \\
  \noalign{\smallskip}
6 & 1873  & 12/07/2002 & 90 & 20000& 2452469.4313  &
$38.1\pm0.7$ & G1 V& +0.5$\pm$0.4 & 7.08\\
&  & 15/02/2003 &    & 20000& 2452685.7243  & $29.4\pm1.0$
& \\
&  & 12/05/2003 &    & 20000& 2452772.4789  & $28.7\pm0.3$
& \\
  \noalign{\smallskip}
7 & 217  & 12/07/2002 & 90 & 20000& 2452469.3571  & $3.8\pm0.7$
&  K4 V& +0.3$\pm$0.3 & 4.87\\
&  & 15/02/2003 &    & 20000& 2452685.4985  & $4.0\pm1.0$
& \\
&  & 12/05/2003 &    & 20000& 2452772.4015  & $3.7\pm0.2$
& \\
  \noalign{\smallskip}
  \hline
  \noalign{\smallskip}
  \end{tabular}
  \end{table*}

\subsection{NGC~5385}
The vector point diagram for all the candidate member
stars having proper motion measurements in the field of NGC~5385
(see Fig.~7 for the identification) from Tycho~2 catalogue is shown
in Fig.~8, where panel a) present the Color-Magnitude Diagram (CMD) derived from
our photometry, whereas panel b) shows the proper motions distribution.
Of the 10 brightest stars which presumably  define the aggregate
(they clearly detach from the bulk of the stars), 7  are
plotted with open squares, and these are the stars for which we obtained
a radial velocity measurement (see below).
However, we found in the Tycho~2 catalogue
proper motion for all these 10 stars, and the 3 for which no radial velocity
measurement is available are plotted with filled triangles. 
The remaining stars are finally plotted with open circles.\\

It is evident that these stars do not share a common mean tangential motion. 
In fact individual proper motion vectors range from
about 5 to 63 mas yr$^{-1}$, while the typical uncertainties of the proper motions
is about 2 mas yr$^{-1}$ per component. This spread in tangential motion
is seen to be in agreement with that shown by field stars 
in the surrounding of NGC~5385. We are therefore inclined to suggest
that the stars belonging to NGC~5385 simply are part of the local
Galactic field.\\

\subsection{NGC~2664}
The vector point diagram for all the candidate member
stars having proper motion measurements 
 in the field of NGC~2664
(see Fig.~9 for the identification) from the Tycho-2 catalogue is shown
in Fig.~10, together with the CMD obtained from our photometry. 
Also in this case, there is a group of stars sharply detached
from the bulk of the stars in the field.
These are 4 stars, and are plotted with squares in both
the panels. \\

It is evident that these stars do not share a common mean tangential motion. 
In fact individual proper motion vectors range from
about 4 to 26 mas yr$^{-1}$, while the typical uncertainties of the proper motions
is about 2 mas yr$^{-1}$ per component. Only, we note that stars $\#$2354 and
$\#$2400 have quite a similar (within the errors) tangential motion, 
and might constitute a physical pair.
The spread in tangential motion
is seen also in this case in agreement with that shown 
by field stars in the surrounding of NGC~2664.

  \begin{table*}
 \tabcolsep 0.40truecm 
  \vspace{1.cm}
  \caption[]{Spectroscopic results for NGC~2664}
  \label{tab2}
  \begin{tabular}{cccccccccc}
  \hline
  \noalign{\smallskip}
  \noalign{\smallskip}
  ID &TYC & Date of& S/N & Resolution&Julian Date  & Rad. vel
& Sp. Type & [Fe/H] &(m-M)$_V$\\
  &816-   & Observation &  &   & & [km/s] &  & [dex]& [mag]\\
  \noalign{\smallskip}
  \hline
  \noalign{\smallskip}
 1 &2354  & 15/01/2003 & 160 & 20000&2452655.5065& $26.0\pm 2.0$ & F5 IV-V& +0.3$\pm$0.3&7.70\\
   &  & 12/05/2003 &     & 20000&2452772.3168& $27.1\pm 0.3$& \\
  \noalign{\smallskip}
 2 & 1826  & 15/01/2003 & 130 & 20000&2452656.4174& $13.5\pm 1.5$ & & & \\
   &  & 06/12/2003 &     & 20000&2452979.5724& $54.4\pm 3.5$& \\
  \noalign{\smallskip}
3 &1890  & 15/01/2003 & 140 & 20000&2452655.5844& $14.4\pm 1.5$ & K0 III& +0.6$\pm$0.4 & 9.79\\
  &  & 12/05/2003 &     & 20000&2452772.3446& $12.7\pm 0.3$& \\
  \noalign{\smallskip}
4 & 2400  & 15/01/2003 & 110 & 20000&2452656.5643& $23.0\pm 1.5$ & K1 III& +0.1$\pm$0.3 & 10.41\\
  &  & 06/12/2003 &     & 3600 &2452979.6014& $20.5\pm 4.1$ & \\
  \noalign{\smallskip}
  \hline
  \noalign{\smallskip}
  \end{tabular}
  \end{table*}

  \begin{table*}
 \tabcolsep 0.40truecm 
  \vspace{2cm}
  \caption[]{Spectroscopic results for Collinder~21}
  \label{tab2}
  \begin{tabular}{ccccccccc}
  \hline
  \noalign{\smallskip}
  \noalign{\smallskip}
  ID & TYC & Date of& S/N & Resolution &Julian Date  & Rad. vel
& Sp. Type & (m-M)$_V$\\
  & 1759-   & Observation &  &  & & [km/s] &  & [mag]\\
  \noalign{\smallskip}
  \hline
  \noalign{\smallskip}
1 & 450  & 04/07/2002 &  530 &20000 &2452468.5756& $-3.2\pm 0.7$
& F9 V & 3.90\\
&  & 20/09/2003 &  300 & 3600 &2452902.5724& $7.4\pm 5.5$
& \\
&  & 05/12/2003 &      & 3600 &2452979.1951& $42.4\pm 4.8$
& \\
  \noalign{\smallskip}
2 & 1472  & 20/09/2003 &  360 & 3600 &2452902.5906& $5.3\pm 6.2$
& G0 V & 5.35\\
&  & 05/12/2003 &      & 3600 &2452979.2083& $11.7\pm 5.7$
& \\
  \noalign{\smallskip}
3 & 1025  & 20/09/2003 &  330 & 3600 &2452902.6222& $55.2\pm
7.8$ & M8 III & 10.84\\
&  & 05/12/2003 &      & 3600 &2452979.4393& $68.3\pm 11.4$
& \\
  \noalign{\smallskip}
4 & 468  & 09/10/2003 &  210 & 3600 &2452922.4483& $-25.5\pm
4.4$ & K2 III & 10.52\\
&  & 05/12/2003 &      & 3600 &2452979.2983& $-15.9\pm 10.3$
& \\
  \noalign{\smallskip}
5 & 994  & 09/10/2003 &  230 & 3600 &2452922.4725& $46.9\pm 7.3$
& G8 V & 5.67\\
&  & 05/12/2003 &      & 3600 &2452979.2378& $37.5\pm 8.0$
& \\
  \noalign{\smallskip}
6 & 462  & 09/10/2003 &  210 & 3600 &2452922.5142& $42.9\pm 7.4$
& G3 V & 6.32\\
&  & 05/12/2003 &      & 3600 &2452979.3997& $24.8\pm 6.4$
& \\
  \noalign{\smallskip}
7 & 1333  & 05/12/2003 &  190 & 3600 &2452979.5238& $48.5\pm
4.6$ & K2 V & 5.31\\
&  & 04/01/2004 &      & 3600 &2453009.3715& $57.0\pm 5.2$
& \\
  \noalign{\smallskip}
8 & 266  & 20/09/2003 &  230 & 3600 &2452902.4968& $50.7\pm 4.6$
& G8 III & 9.86\\
&  & 04/01/2004 &      & 3600 &2453009.3979& $40.4\pm 5.2$
& \\
  \noalign{\smallskip}
9 & 866  & 20/09/2003 &  310 & 3600 &2452902.4559& $3.5\pm 5.7$
& G5 III & 9.49\\
&  & 04/01/2004 &      & 3600 &2453009.4311& $17.2\pm 7.9$
& \\
  \noalign{\smallskip}
10 & 1114  & 20/09/2003 &  270& 3600 &2452902.5238& $48.5\pm 4.6$
& K1 III & 9.36\\
&  & 04/01/2004 &     & 3600 &2453010.3701& $5.2\pm 5.7$
& \\
  \noalign{\smallskip}
11 & 1749  & 20/09/2003 &  150& 3600 &2452980.2946& $24.8\pm 7.2$
& F8 V & 8.14\\
&  & 04/01/2004 &     & 3600 &2453010.3303& $23.0\pm 7.1$
& \\
  \noalign{\smallskip}
12 & 272  & 20/09/2003 &  120& 3600 &2452980.3317& $33.9\pm 7.3$
& G0 V & 8.36\\
&  & 04/01/2004 &     & 3600 &2453010.2885& $41.2\pm 8.3$
& \\
  \noalign{\smallskip}
  \hline
  \noalign{\smallskip}
  \end{tabular}
  \end{table*}

\noindent

\subsection{Collinder~21}
The CMD and the vector point diagram for all the candidate member
stars having proper motion measurements in the field of Collinder~21
(see Fig.~11 for the identification) from the Tycho-2 catalogue are shown
in Fig.~12 panels. 
12 bright stars, in the magnitude range $8 \leq V \leq 13$, define the aggregate
and are plotted in both panels as open squares.
Proper motions are also available for 3 dimmer stars (solid triangles)
which bridge (see the CMD in panel a) ) the brightest
group to  the bulk of the stars in the field. However these 3 stars have
significantly
diverse proper motion components (see Table~5). 
Coming back to the 12 brightest stars, their proper motion vectors range from 
about 3 to 68 mas yr$^{-1}$, while the typical uncertainties of the proper motions
is about 2 mas yr$^{-1}$ per component.
It is evident that these stars do not share a common mean tangential motion. 
There are two clear pairs. 
One is the well known visual binary system BD+26~305ab (stars
$\#$866 and $\#$1114), and the other is  composed by the stars $\#$1472 and $\#$272,
which however
deserve further investigation.
Apart from these two pairs, the bulk of the stars exhibit a disordered motion.

\section{Spectroscopic analysis}
Multi-epoch spectra have been acquired for all the candidate 
members (23 stars in total for a grand total of 54 spectra) in the aggregates
under investigations. The results of the spectroscopic survey are listed in
Tables 6 to 8.

\subsection{NGC~5385}
For this object we provide 3 epochs Echelle spectra for seven bright stars,
which
turn out to be all dwarfs (see Table~6), but for star $\#$1797, which seems
to be a subgiant.
By looking at this table results
we find that 2 stars are probably unresolved binary stars ($\#$1641 and $\#$1873), whereas
all the other stars
do not exhibit significant variations in their radial velocity. 
The studied stars have radial velocity ranging from -62 km/s to +30 km/s, and
they
differ one from the other, thus suggesting that NGC~5385 is not a physical
aggregate.
There might be some common motion pairs in this sample. Stars $\#$399 
and $\#$1797
(see Table~3) have marginally consistent proper motion components,
and stars $\#$1641 and $\#$217 as well. The first couple has also compatible
spectral type (see Table~6). Radial velocities however (see Table~6)
 contradict this hypotesis. Moreover both pair members are located 
in very different posistion (see Fig.~7).\\
We derived stellar distances from the Sun basing on the spectral classification
reported in Table~6.  The relation between spectral type, luminosity class and
absolute magnitude has been taken from the Michigan Spectral Catalogue
Project.\footnote{http://www.astro.lsa.umich.edu/users/hdproj/mosaicinfo/absmag.html.}
\noindent
Distance moduli range from 4.87 to 8.17 mag, which turn
into distances going from 94 to 430 pc. Schlegel et al (1998) extinction
maps provide for NGC 5385 region a reddening of 0.04 mag, which does not signifcantly
alter the derived distances. The significant spread of stars distances 
rules out the possibility they form a physical group.\\
Metal abundances were derive using
the MOOG code (freely distributed by Chris Sneden, University 
of Texas, Austin) as described in Carraro et al. (2004).
The $S/N$  ratio and the limited resolution prevented us from deriving
very precise metallicity estimates, and therefore the value reported in Table~6
must be taken as an indication. We find that within the uncertainties
the bulk of the stars possess solar iron abundance, but for star $\#$457,
which exhibits a super solar iron abundance. 
Within 3$\sigma$ the metallicity distribution looks homogeneous (0.4$\pm$0.2),
although we believe that more precise estimates of the metallicity would probably 
reveal some spread, which is expected for a random group of stars.

\subsection{NGC~2664}
For this object we provide 2 epochs Echelle spectra. From the analysis of
Table~7 results
we find that one star is a spectroscopic binary ($\#$1826), 
and due to blending problems we were not able to derive
either stars spectral type or metallicity.
The remaining three stars
do not exhibit significant variations in their radial velocity.
Star $\#$2354 is a  dwarf, while stars  $\#$1890 and $\#$2400 are  giants.
The studied stars have radial velocity ranging from +12 km/s to +30 km/s, and
they
basically differ one from the other, thus suggesting that NGC~2664 is not a 
physical aggregate.
The proper motion pair ($\#$2354 and $\#$2400) seem to be actually a pair, since
the radial velocities of the two stars do not differ very much (see Table~7).\\
As for NGC~5385, we computed spectral type based distances, which are
listed in the last column of Table~7. The distance moduli range
from 7.70 to  10.41 mag., which implies distances ranging from 350 to 1207 pc,
which slight corrections due to the interstellar reddening, which
in the direction of NGC~2664 amounts to 0.02 mag (Schlegel et al. 1998).
The four stars which define the group cannot be considered physically bound.
Finally, the possible related stars $\#$2354 and $\#$2400  lie too apart
to be considered as a physical pair.\\  
In the same table we list estimates
of the metal content of  three stars, which turns out to be around solar
or slighly supersolar for all the stars within the errors.

\subsection{Collinder~21}
For this object we provide 2 epochs medium resolution spectra for all stars but
for
$\#$450, for which we add an Echelle spectrum.
5 stars turn out to be giants, and all the other are dwarfs. 
Though in this case the errors are much larger, from 
the analysis of Table~8 results we confirm that this star ($\#$450) is a binary.
Large velocity variations are also shown by the couple $\#$866-$\#$1114
(BD+26~305a,b),
which however we are inclined to ascribe to their nature of visual binary.
All the other stars show deviating radial velocity, roughly ranging from
-20 km/s to +60 km/s,
thus suggesting that this circlet  is not a physical aggregate. The possible proper
motion pair ($\#$1472-$\#$272) is not confirmed by radial velocity measurements,
and, on top of that, these two stars are located in quite different places
(see Fig.~11).\\
As for the previous objects we have derived stars individual distances based
on spectral classification. The distance moduli range from  3.9 to 10.84 mag,
which means a distance between 60 and 1470 pc. The reddening in this direction
is 0.07 mag (Schlegel et al. 1998), and therefore these estimates do not
change significantly due to it. 
The distance spread is too huge to be consistent with Collinder 21
being a physical aggregate.\\
\noindent
Unfortunately, in this case we do not have
sufficiently high resolution spectra to derive reasonable estimates
of the metal abundance.

\begin{figure}
\includegraphics[width=9cm]{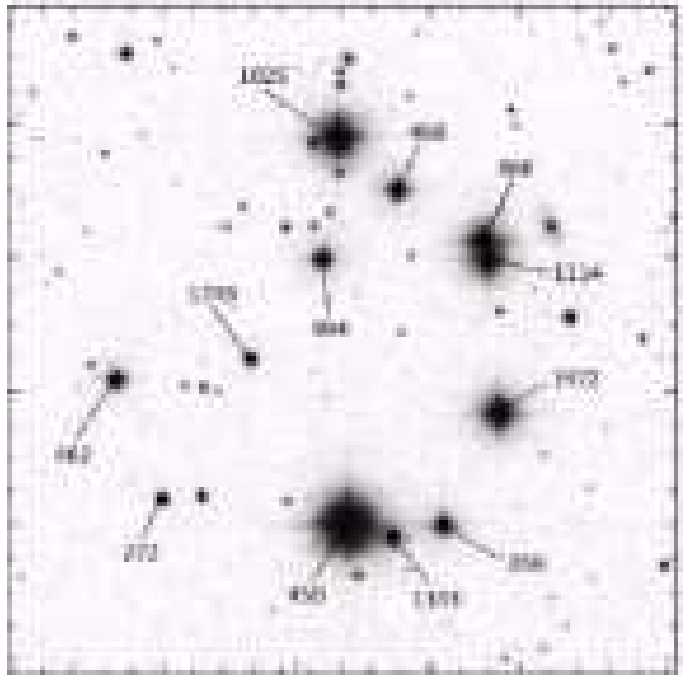}
\caption{Identification map ($8^{\prime}.14 \times 8^{\prime}.14$, taken from
DSS-2)
for member candidates of Collinder~21 having proper motions measurements from
Tycho~2. North is up, East on the left. The labels give the star numbers
assigned in
the Tycho-2 catalog (in the sense TYC 1759$-<...>$).
Data for these stars
are given
in Tables~5 and 8.}
\end{figure}

\begin{figure*}
\includegraphics[width=18cm]{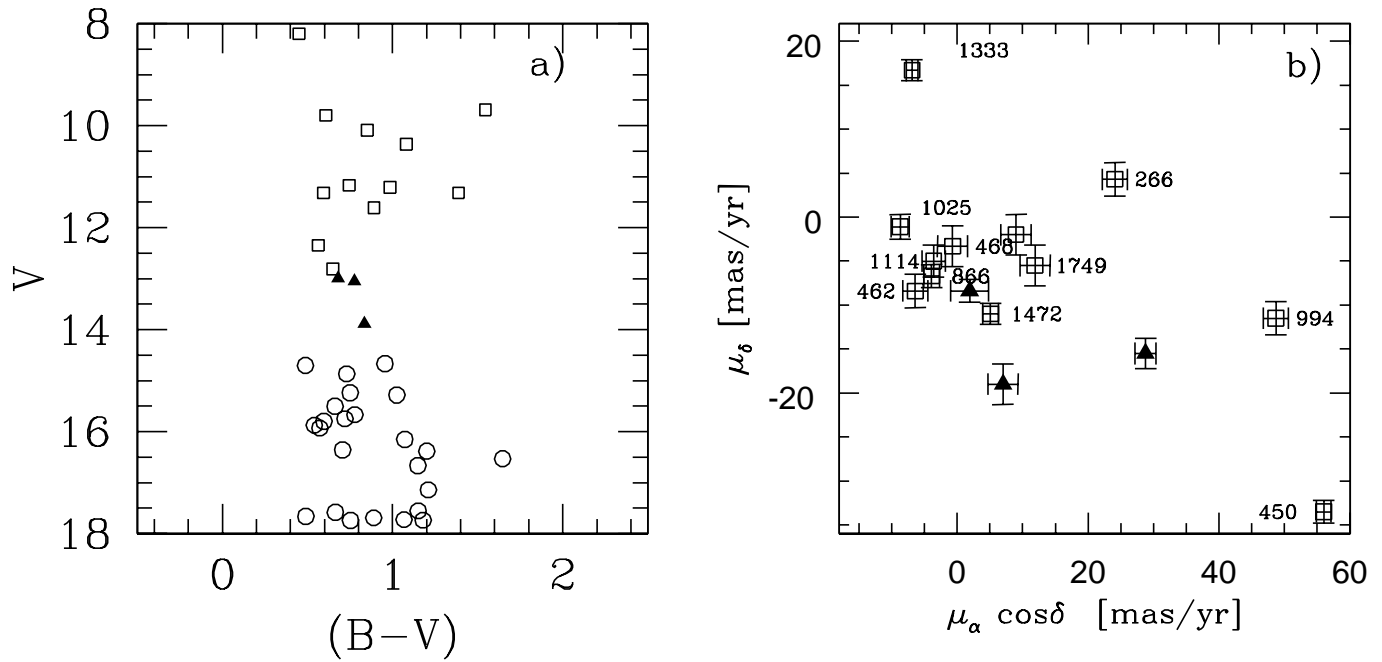}
\caption{Photometry and proper motions in the region of Collinder~21.
{\bf a)} Color-Magnitude diagram. Open squares are Tycho~2 stars 
having both proper motion and radial velocity measurements
(see panel b) and Table~8), while
filled triangles are stars having only proper motion
measurements. {\bf b)} Vector point
plot of Tycho~2 proper motions and proper motion errors. 
Symbols are as in panel {\bf a)}.As for the identification,
we are plotting here Tycho~2 numbering. See text for more details.}
\end{figure*}

\section{Discussion and Conclusions}
In this paper  we have analysed the possibility that NGC~5285, NGC~2664 and
Collinder~21
might be POCRs. Our study however yields negative answers for all of them.\\
In fact, star counts and the appearance of the CMDs confirm that we are facing
significant stellar overdensities above the general Galactic field.\\
However, the close scrutiny of the kinematic data (proper motion and radial
velocity)
provide us with a completely diverse picture.
All the agglomerates are not physical clusters since they show very different
velocity
components, and therefore they force us to consider them simply as projective
chance
alignment of unrelated field stars.\\
We found out a number of possible unresolved binary stars or common motion
pairs which might deserve further investigation.\\
The probale binary fraction we found ranges from about 25$\%$ in
 NGC~2664 and NGC~5385 
to 18$\%$  in Collinder~21, and it is not much different from the typical
Galactic disk field binary percentage.\\
\noindent
Our findings confirm previous suggestions (Odenkirchen \& Soubiran 2002) that
an overdensity
of stars does not necessarily 
imply the existence of a physical ensemble, and cast some doubts
on the POCRs list provided by Bica et al. (2001), which is purely based
on star counts. 
Nevertheless, work is in progress to define a more efficient OCRs
finding criterion (de la Fuente Marcos, in preparation)
and to probe the real nature of other POCRs candidates.

\begin{acknowledgements}
The entire Asiago technical staff is deeply acknowledged 
for the kind night assistance over the whole duration of
this project. GC thanks Brian A. Skiff for providing
continuous and very useful comments.
This study made use of Simbad and WEBDA. 
 \end{acknowledgements}


\begin{thebibliography}{}
\bibitem{} Bassino, L.P., Waldhausen, N. \& Mart\'{\i}nez, R.E. 2000, 
              A\&A, 355, 138 
\bibitem{} Baume, G., Villanova, S., Carraro, G., 2003, A\&A 407, 527
\bibitem{} Baumgardt, H. 1998, A\&A, 340, 402
\bibitem{} Bica, E., Santiago, B.X., Dutra, C.M., et al. 2001,
              A\&A, 366, 827 
\bibitem{} Carraro, G. 2000, A\&A, 357, 145
\bibitem{} Carraro, G. 2002, A\&A, 385, 471
\bibitem{} Carraro, G., Ng, Y.K., Portinari, L. 1998, MNRAS, 246, 1045
\bibitem{} Carraro, G., Bresolin, F., Villanova. S., et al. 2004,
           AJ, in press ({\tt astro-ph/0406679})
\bibitem{} de la Fuente Marcos, R. 1997, A\&A, 322, 764
\bibitem{} de la Fuente Marcos, R. 1998, A\&A, 333, L27
\bibitem{} Desidera, S., Fantinel, D., Giro, E. 2001, AFOSC USER MANUAL
\bibitem{} Dias W.S., Alessi B.S., Moitinho A. \& Lepine J.R.D.,  2002, A\&A, 389, 871
\bibitem{} Halbwachs, J.L., Mayor, M., Udry, S., Arenou, F., 2003, A\&A 397, 159 
\bibitem{} H{\o}g, E., Fabricius, C., Makarov, V.V., et al. 
              2000, A\&A, 355, L27
\bibitem{} Landolt, A.U. 1992, AJ, 104, 340
\bibitem{} Monet D.G., Levine S.E., Casian B., et al. 2003, AJ, 125, 984
\bibitem{} Odenkirchen, M., Soubiran, C. 2002, A\&A, 383, 163
\bibitem{} Patat, F., Carraro, G. 2001, MNRAS, 325, 1591
\bibitem{} Pavani, D.B., Bica, E., Dutra, C.M., et al. 2001, 
              A\&A, 374, 554
\bibitem{} Schlegel, D.J., Finkbeiner, D.P., Davis, M. 1998, ApJ, 500, 525
\bibitem{} Villanova, S. 2003a, Master Thesis, Padova University
\bibitem{} Villanova, S., Carraro, G., de la Fuente Marcos, R. 2003b, in 
           "Milky Way Surveys: Structure and Evolution of our Galaxy",
           proceedings of the 5th Boston University Astrophysics Conference, in
press
\bibitem{} Villanova, S., Baume, G., Carraro, G. \& Geminale, A. 2004, A\&A 419, 149
\bibitem{} Zacharias, N., Urban, S.E., Zacharias, M.I., et al. 2003, AJ, in
preparation
 \end{thebibliography}
 \end{document}